\documentstyle[12pt,equation]{article}
\advance\voffset by 28 pt
\begin{document}
\newcommand{\ccbar}{\mbox{$c \overline{c} $}}
\newcommand{\Y}{\mbox{$\Upsilon$}}
\newcommand{\aem}{\mbox{$\alpha_{{\rm em}}$}}
\newcommand{\rmcosh}{\mbox{\rm cosh}}
\newcommand{\rmln}{\mbox{\rm ln}}
\newcommand{\rmgev}{\mbox{\rm GeV}}
\newcommand{\shat}{\mbox{$\hat{s}$}}
\newcommand{\that}{\mbox{$\hat{t}$}}
\newcommand{\uhat}{\mbox{$\hat{u}$}}
\newcommand{\sighat}{\mbox{$\hat{\sigma}$}}
\newcommand{\rs}{\mbox{$\sqrt{s}$}}
\newcommand{\pT}{\mbox{$p_T$}}
\newcommand{\xq}{\mbox{$(x,Q^2)$}}
\newcommand{\qi}{\mbox{$q_i (x,Q^2)$}}
\newcommand{\ggl}{\mbox{$g (x,Q^2)$}}
\newcommand{\qiA}{\mbox{$q_i^{A}$}}
\newcommand{\gA}{\mbox{$g^{A}$}}
\newcommand{\dsgmdpt}{\mbox{${d\sigma}/{dp_T}$}}
\newcommand{\gp}{\mbox{$g^p$}}
\newcommand{\rhog}{\mbox{$\rho_g $}}
\newcommand{\xone}{\mbox{$x_1$}}
\newcommand{\xtwo}{\mbox{$x_2$}}
\newcommand{\qqbar}{\mbox{$q \overline{q}$}}
\newcommand{\ppbar}{\mbox{$p \overline{p}$}}
\newcommand{\jpsi}{\mbox{$J/\psi$}}
\newcommand{\xt}{\mbox{$ x_T\ $}}
\newcommand{\xbt}{\mbox{$\bar x_T\ $}}
\newcommand{\yj}{\mbox{$ y^{J/\psi}$}}
\newcommand{\ygam}{\mbox{$y^{\gamma}$}}

\newcommand{\be}{\begin{equation}}
\newcommand{\ee}{\end{equation}}
\newcommand{\een}{\end{subequations}}
\newcommand{\ben}{\begin{subequations}}
\newcommand{\beq}{\begin{eqalignno}}
\newcommand{\eeq}{\end{eqalignno}}
\renewcommand{\thefootnote}{\fnsymbol{footnote} }
\pagestyle{empty}
\noindent
{\flushright CERN-TH.6679/92\\
TIFR/TH/92-59\\
BU-TH-92/3\\}
\vspace{2cm}
\begin{center}
{\Large \bf Looking for the gluonic EMC effect in associated\\
$J/\psi$ + $\gamma $ production}\\
\vspace{5mm}
{R.V. Gavai$^{1,a)}$, R.M. Godbole$^{2,b)}$ and K. Sridhar$^{3,c)}$}\\
\vspace{5mm}
{\em 1. Theory Group, Tata Institute of Fundamental Research,
Bombay 400005, India} \\
{\em 2. Dept. of Physics, Univ. of Bombay, Vidyanagari, Bombay
400098, India}
\\
{\em 3. Theory Division, CERN, CH-1211, Geneva 23, Switzerland.}\\
\end{center}

\vspace{2.cm}
\begin{abstract}
We study the associated production of $ J/\psi\ + \gamma $ in
fixed-target experiments with nuclear targets as well as at the
relativistic heavy-ion collider (RHIC). We find that this process
affords a very clean probe of $\rho_g$, the ratio of gluon density in a
heavy nucleus to that of a proton.  The combined $x$ range thus available
can be used to discriminate between the predictions
of different models of the EMC effect for $\rho_g$.

\end{abstract}

\vspace{3cm}
\noindent
$^{a)} $ gavai\%theory@tifrvax\\
$^{b)} $ rohini\%theory@tifrvax\\
$^{c)} $ sridhar@vxcern.cern.ch\\

\vspace{36pt}
\noindent
CERN-TH.6679/92\\
October 1992\\
\vfill
\clearpage
\setcounter{page}{1}
\pagestyle{plain}

The importance of a good knowledge of the parton densities in
hadrons cannot be overemphasized in view of their significance
for the physics that can be studied at
both the \ppbar\ and the relativistic heavy ion colliders. The
parton distributions, which are often obtained by using nuclear targets,
have a non-trivial dependence on the environment, viz. the well-known EMC
effect \cite{emcdata,emcrev}. While the nuclear effects on the quark densities
can be directly studied in deep inelastic scattering
(DIS) experiments, the gluon
densities \ggl\ get determined only indirectly, through their effect
on the QCD evolution of the structure function. A good determination of
the gluon densities in nuclei is possible only through studies of
different hard scattering processes,  dominated by gluons, in
photon/lepton/hadron - nucleus collisions. A direct determination of the
nuclear gluon densities will help in pinning down the correct model of the
EMC effect. The theoretical basis of this effect still remains unclear
as there are a large number of models \cite{emcrev}, all of which are in
reasonable conformity with the data available from DIS.  However,
these models give very distinct predictions for \rhog\ defined as,
\be
\label{e1}
\rho_g = g^A(x,Q^2)/g^p(x,Q^2) .
\ee
Thus experimental information on this ratio may help in clarifying the
dynamical origin of the EMC effect.

The standard processes that have been used to measure gluon densities
in protons are large-\pT\ jet production, heavy flavour production
and direct photon/dimuon production at large $p_T$ in hadron-nucleon
collisions, and lepto- and photoproduction of quarkonia with nucleon
targets.The same processes can, of course, be used as an effective tool
to determine the gluonic EMC ratio, \rhog\ \cite{adep,sousri}.
In fact the NMC collaboration \cite{nmc} has measured the ratio
of gluon distributions in Sn and C, by studying inelastic $J/\psi$
production by scattering 200 and 280~GeV muon beams off these
targets. With the cuts imposed in this experiment, the
values of $x$ probed are in the range 0.05 to 0.15. It
will be interesting to complement this information with gluon
distributions from other processes mentioned above.
With the exception of lepto- and photoproduction
of $J/\psi$, and jet production at lower \pT\ values,
all the other processes receive contributions from quarks
as well. So one has to either choose special kinematic regions or
find clever combinations of different cross-sections \cite{sousri}
to separate the quark contributions from those of the gluon.

In the context of the colour singlet model \cite{csmod}, it has been shown
recently \cite{drekim} that a study of the
associated production of \jpsi\ and $\gamma$ in hadronic collisions
affords a very clean determination of the gluon density. In this
process, one looks for a photon recoiling against the \jpsi\ produced
at large $p_T$.
The authors of Ref.~\cite{drekim} have studied this process
for the \ppbar\ and the HERA $ep$ collider and their study
has been later extended \cite{sri2}  to polarised $pp$ scattering.
In this note, we analyse the effectiveness
of this process to probe the gluon densities in nuclei in fixed-
target experiments and at the RHIC.
We find that, owing to the lower values of \rs\
in fixed target experiments, the range of $x$ values for which
\gA\ can be probed increases considerably.
This can be done by studying the triple differential cross-section $d\sigma/
{dp_T\ dy^\gamma dy^{J/\psi}}$ for negative rapidities of the
photon and \jpsi.  Moreover, when the \jpsi\ is tagged through its
leptonic decay the final state is particularly clean. This further
increases the efficacy of this process in probing \gA\ at larger
values of $x$, as compared with direct photon production \cite{sousri}.
The study of this process at the RHIC will be shown to probe \gA\ at the
same small values of  $x$  as in lepto- and photoproduction of \jpsi.
Determination of $\rho_g$ using this process turns out to have two
additional advantages: i) it is less sensitive to the theoretical
uncertainties of the production mechanism, and ii) it is directly
proportional to the ratio of experimentally measured cross-sections.

Since some of the statements regarding the production process
depend crucially on the hadronization mechanism in the production
of \jpsi,  it is worthwhile summarising its main features.
\jpsi\ production is assumed to take place through
the mechanism specified in the colour singlet model \cite{csmod}.
In this model, one projects out the state with the correct spin,
parity and colour assignments from the full \ccbar\ production
amplitude, to match the quantum numbers of the $J/\psi$. This
projection is done at the level of the hard scattering amplitude
itself, which yields a multiplicative factor, $R(0)$, the \jpsi\
wavefunction at the origin in coordinate space.

The basic subprocess, at the parton level, which is responsible
for the final state with the \jpsi\ and the photon at large \pT\ is
\be
\label{e2}
g + g  \rightarrow J/\psi + \gamma .
\ee
The \qqbar\ initial state does not contribute because in that case
the produced \ccbar\ pair is not a colour singlet. Thus this final
state is accessible only to gluons, and therefore depends directly
on the gluon distribution. Moreover, the $P$-state charmonia do not
get produced in this process. This is so because the subprocess
\be
\label{e3}
g + g \rightarrow \chi \rightarrow J/\psi +\gamma
\ee
does not produce the $J/\psi$ and photon at large \pT, whereas the
subprocess
\be
\label{e4}
g + g \rightarrow \chi + \gamma
\ee
is disallowed because of $C$-invariance and colour neutrality of the final
state. Thus, this process is purely gluon-initiated and one does not have
to worry about $\chi$ production and the subsequent decay of the $\chi$
into a $J/\psi$. It is worth noting that for the $P$-states it is
the derivative of the bound-state wavefunction at the origin, and not the
wavefunction itself, that is relevant. From simply this consideration,
one expects $\chi$ production to be suppressed by a factor of 50, given
that the amplitudes for the processes in eqs.~(\ref{e2}) and (\ref{e4})
are of the same order of magnitude. Also, outside the framework
of the colour singlet model the \qqbar\ initial state can give rise to a
final state containing a quarkonium and $\gamma$. However, this production
mechanism will give rise to additional hadronic activity in the vicinity
of the photon.  Hence demanding a high-\pT , isolated $\gamma$ will always
ensure that the process is gluon-initiated.

It may be noted that in the framework of the  colour singlet model the
formation of \jpsi\ takes place on a perturbative time scale, being $\sim
1/m_{J/\psi}$. This can have extremely interesting implications for the
\jpsi\ suppression signal \cite{satz} of quark gluon plasma formation,
which may also have
a comparable or even larger formation time.  Thus the only effect the
plasma can have in this case will be on the propagation of the \jpsi.
This makes the physics of confinement/deconfinement cease to have any
significant effect on the experimentally observed suppression of the
\jpsi\ cross-sections.  Instead the physics of its relative propagation
through quark gluon plasma and a hadronic gas assumes an important role,
bringing out further the model-dependence of the proposed signal.

\vspace{4pt}
The colour singlet model detailed above has been used in the past to
describe both leptoproduction and hadroproduction of $J/\psi$,
and is known to give a good description of the
kinematical distributions \cite{nmc,uaone}. However, it has been
found that there is considerable uncertainty in the overall
normalisation for \jpsi\ production. This is probably due to
the non-relativistic treatment of the \jpsi\ in this model, as
well as to the small mass of the charm quark. However,
we propose to study ratios of the cross-sections for different
targets and hence the effect of these K-factors will be reduced.
Moreover, the data \cite{nmc} require a K-factor $\ge 1.0 $.
This would mean that our cross-sections  can indeed be considered
as a conservative estimate.

\vspace{4pt}
The cross-section for the subprocess $g+g \rightarrow J/\psi +
\gamma$ is given in the colour singlet model as
\cite{csmod,drekim}
\beq
\label{e5}
{d\sighat \over d \that} = &{16\pi \alpha \alpha_s^2 M_{\psi}
\vert R(0) \vert^2 \over 27\shat^2} \biggl\lbrack {\shat^2
\over (\that -M_{\psi}^2)^2(\uhat -M_{\psi}^2)^2} \cr
& +{\that^2 \over (\shat - M_{\psi}^2)^2(\uhat -M_{\psi}^2)^2}
 +{\uhat^2 \over (\that - M_{\psi}^2)^2(\shat -M_{\psi}^2)^2}
\biggr\rbrack ,
\eeq
where \shat, \that\ and \uhat\ are the Mandelstam variables for
the subprocess, and the modulus squared, $\vert R(0) \vert^2$,
of the wavefunction is related to the leptonic decay width by
\be
\label{e6}
\vert R(0) \vert^2 = {9 M_{\psi}^2 \over 16 \alpha^2} \Gamma
(J/\psi \rightarrow e^+ e^-) = 0.544\ {\rmgev}{}^3 ,
\ee
if the value of the leptonic width of the \jpsi\ is taken
to be 5.36~keV.

\vspace{4pt}
The triple differential cross-section
$d\sigma /{ dp_T\ dy^\gamma dy^{J/\psi}} $ for the reaction
$B + A \rightarrow J/\psi + \gamma +X$  is then given by
\be
\label{e7}
{{d^3 \sigma} \over {d\yj d\ygam  dp_T }}  = 2 p_T x_1 x_2 g^B (x_1)
                     g^A (x_2) {{d \sighat} \over {d \hat t}} ,
\ee
where $g^{B}(x_1)$ and $g^{A}(x_2)$ are the gluon distributions
in the beam $B$ and the nuclear target $A$ respectively,
$ {{d \sighat} /{d \hat t}}$ is the subprocess cross-section given
by eq. (\ref{e5}) and $y^{J/\psi}$ and $y^{\gamma}$ are the rapidities,
in the $pA$ centre-of-mass frame, of the \jpsi\ and the photon
respectively. The momentum fractions \xone\ and \xtwo\ of the beam and
the target carried by the gluons are given by the following kinematic
relations:
\beq
\label{e8}
x_1 = &{1 \over 2} \lbrack \bar x_T e^{y^{J/\psi}}
+ x_T e^{y^{\gamma}} \rbrack , \cr
x_2 = &{1 \over 2} \lbrack \bar x_T e^{-y^{J/\psi}}
+ x_T e^{-y^{\gamma}} \rbrack .
\eeq
In the above equation  $x_T = 2p_T/\sqrt{s}$ and
$\bar x_T = \sqrt{x_T^2+4\tau}$ with
$\tau = M_{\psi}^2/s$. Note that since in this process, the \jpsi\
and the photon are both detected, the values of $p_T$,
$y^{J/\psi}$ and $y^{\gamma}$ are all known and therefore the
kinematics is determined completely. The measured cross-section
is, therefore, fully differential in the kinematic variables.

We will also discuss the partially integrated cross-section
${d\sigma} / {dp_T}$, which is obtained by integrating the
differential cross-section given in eq. (\ref{e7}) over
restricted regions of
phase space. For a given value of $x_T$ the allowed range of
phase space is given by
\beq
\label{e9}
|y^{J/\psi}| \leq ~&{\rmcosh}^{-1} \bigg[{{(1 + \tau )} \over {\xbt}}
\bigg ], \cr
-{\rmln} \big({{2 - \xbt e^{-\yj}} \over \xt  } \big) &\leq \ygam \leq
{\rmln}   \big({{2 - \xbt e^{\yj}} \over  \xbt} \big).
\eeq

In order to
illustrate the effectiveness of this process in differentiating between
different models of the EMC effect, we have chosen three representative
cases. These are a) the gas model \cite{gas}, b) the rescaling model
\cite{rescale}, and c) the six-quark cluster model \cite{bag}.  As was
already mentioned, in all these models  the construction of nuclear
parton densities starts from a choice of nucleon parton densities. The
model parameters are then fixed, using the DIS data with nuclear targets.
We refer the reader to the original papers for the details of the
models and their fits to the DIS data.  The model (b) has no cumulative
effects whereas the others are two-component models where the nuclear
structure function has two contributions -- one from the normal nucleons
and the other from nucleons that share their partons with the other
nucleons.  In Fig. 1  we plot the ratio \rhog\ defined in eq.(\ref{e1}) as
a function of \xtwo\ for the three models for $A=56$\footnote{A
similar figure appears in the paper by Godbole and Sridhar
(Fig.~1 of that paper) quoted in Ref.~\cite{adep}. The curves
for the gas model and the rescaling model shown there are incorrect
and should be replaced with those shown here.}.
It may be noted
here, however, that \rhog\ does not depend very strongly on the mass
number $A$ of the nuclear target. All the models give somewhat similar
predictions, in the vicinity of 1.0, for
\rhog\ in the range $0.1 \le x_2 \le 0.2 $. The deviation of
the ratio from unity is significant for all the models at larger values of
\xtwo\ and also at the smaller values for the rescaling model.
It should be noted here that the choice of nucleon parton densities in
each model is different. As a result, the relative behaviour of \gA\ in
different models can be different from that for \rhog\ shown here.
The axis on the top of the figure shows, as an illustration, \pT\ values for
the symmetric configuration $y^\gamma = 0 , y^{J/\psi} = 0 $ for the FNAL
beam energy of 800 GeV in the fixed-target mode, i.e., $\sqrt{s} = 38.75$
GeV.  As is clear from eq. (\ref{e7}) the ratio of the triple
differential cross-sections for $pA$ and $pp$, is directly proportional to
\rhog.  Hence this figure also shows the expected ratio of the
cross-sections as a function of \pT\ (for the fixed-target case) for the
symmetric point that we have chosen. Moving away from the symmetric
configuration in the rapidity space to negative values, it is possible
to map large \xtwo\ values to lower \pT\ values than shown in Fig. 1,
where the cross-section is still appreciable.
Conversely, it is possible to probe the smaller values of
$x_2$ by going to positive rapidities. In this way one can obtain
information on gluon densities over a fairly large range of $x$ values.

In Fig. 2 we first show the cross-section \dsgmdpt, for the process
$p + A \rightarrow J/\psi + \gamma \rightarrow l^+ l^- \gamma$ ,
obtained by integrating eq. (\ref{e7}) over the region $|y^{J/\psi,\gamma}|
\leq y_{max}$, for the three models for the fixed-target mode of FNAL,
i.e. $\sqrt{s} = 38.75 $ GeV, and also for $AA$ collisions for RHIC
for $\sqrt{s} = 100$ GeV. We have chosen $y_{max} = 2.0 $ for FNAL and
$y_{max} = 4.0$ for RHIC. Including both electrons and muons will
enhance the cross-section by a factor of 2. We see that, since the
$x_2$-dependence of \gA\ is different in different models, the
$p_T$-dependence of the cross-sections in these models is also quite
different. Recall that the assumed nucleon gluon density in
different models is different. Therefore the gas model predicts
a harder \pT\ distribution than the quark-cluster model, contrary to what one
would expect from Fig. 1. Interestingly, even the lowest prediction for the
FNAL energy, (that of the rescaling model) corresponds to about 30-40 pb
at $p_T = 3 $ GeV.  With the total integrated luminosity of 100 pb${}^{-1}$
that is foreseen in this case, this cross-section corresponds to large
enough statistics to facilitate a good study of the process. Even for $pp$
collisions we should have $\sim 100$ events at $p_T = 3 $ GeV.

The increased  value of \rs\ at RHIC causes the cross-section to be
dominated by the low $x$ region. The \pT\ distribution
is much broader, of course. The much higher cross-sections in this case
compensate for the decrease in luminosity by about a factor of 100 in
going from $pp$ to $AA$ collisions. Since the \gA\ distribution is
much broader for the gas model, the increase in going from  FNAL energies
to RHIC is only a modest factor of 5 (after taking into account the
additional factor of $A$ for the RHIC case). For the other two models the
cross-section increases by more than an order of magnitude. As a matter of
fact the relative size of the cross-sections predicted in the models
changes completely as one goes from FNAL to RHIC. At RHIC
largest (smallest) cross-sections are foreseen for the rescaling (gas) model,
which is exactly opposite to the situation at FNAL. Thus combining
the information on \jpsi\ $+\ \gamma $ production in the fixed target
experiments and at RHIC, one can probe \gA\ over a wide range of
\xtwo\ and can indeed discriminate between different models.  Of course,
owing to the completely differential nature of the measured
cross-section information can be obtained on $\rho_g(x_2) $ over a wide range
of
\xtwo\, even at a fixed value of \rs, by studying the triple differential
cross-sections of eq. (\ref{e7}). We choose $p_T=3 $ GeV for doing this:
at these large values of $p_T$, the prompt photon detection is free of
background and the rates are still appreciable. The decay leptons in this
case will also have on the average $p_T \simeq 2$  GeV.

In Fig. 3 we show the triple differential cross-section for the process
for $A=56$, for the gas model as a function of \yj\ and \ygam , for $p_T = 3$
GeV at FNAL energy.  We have restricted ourselves to values of \yj\
and \ygam\
in experimentally feasible ranges, $|y^{J/\psi},y^\gamma|~ \le 2.0$,
even though the kinematic limit is somewhat higher. As is clear
from the figure, most of the
cross-section is contained in this region. Indeed, the figure
also shows that fairly large values of cross-sections are possible at
large negative and positive rapidities.  These curves show a certain
asymmetry between
\yj\ and \ygam . This is a result of the fact that the nuclear gluon
density is much harder than the nucleon gluon density in this model. For
the quark-cluster model, a very similar configuration results. Only the
absolute values of the cross-section are somewhat lower. The rescaling
model predicts the least cross-section of the three models. This
is a reflection of the softer nuclear gluon density in the
rescaling model and the somewhat large value of \pT\ chosen here. The
softer gluon also results in narrower rapidity distributions in this case.
Thus one sees that the differences in the \gA\ are indeed reflected in
these distributions.

To get a clearer picture of the differences in the predictions between
different models, as well as to gauge the measurability of the process,
we show in Figs.~4(a), (b) and (c) the contours of constant triple
differential cross-sections (in pb/GeV) in the $y^{J/\psi} - y^{\gamma}$
plane, again for
$p_T = 3$~GeV and for the fixed-target case, for the gas model, the
rescaling model and the quark-cluster model respectively. The asymmetry
between \yj\ and \ygam\ is significantly different in all three cases.
The gas model predicts measurable cross-sections ($\sim$~1 pb) all
the way up to $|y^{J/\psi}, y^\gamma| \sim 1.6$, whereas in the rescaling
model these are obtained only up to rather small rapidities of 0.8 or so.
As was said before, this reflects the much softer \gA\ in this case.

Superimposed on these contours are the contours of constant \xtwo . Since
different pairs of \yj, \ygam\ correspond to the same \xtwo\ at a given
\pT,  the ratios of triple differential cross-sections integrated along
this contour will yield \rhog\ at a fixed value of \xtwo.  Note that
while taking the ratios, factors of \xone\ and \pT\ in eq. (\ref{e7}) cancel;
we can thus integrate the ratio along the \xtwo\ contour.  This will
clearly help increase the statistics and reduce the errors without losing
the differential nature of the information.  This enhancement of the
statistics is particularly relevant for proton targets, since it is
necessary to take the ratios of triple differential cross-sections
for the $pA$ and $pp$ case
to obtain $\rho_g$.  From Fig. 4(a), for instance, we see that with
the assumed luminosity of 100 pb${}^{-1}$ at FNAL, for the gas model
we expect $\sim  600 $ events corresponding to $x_2 = 0.3 $. With the
proton (deuterium) target we expect $\sim 15 (30)$ events. A further such
integration is possible along the \pT\ axis as well, although the
cross-section falls rather rapidly in $p_T$, as shown in Fig. 2. Note
that for an effective study of \rhog\ it is necessary that we choose the
lighter target to be proton or deuterium, as the EMC effect saturates
very fast with increasing $A$.

In conclusion we see that the associated production of $J/\psi$ and photon
at large \pT\ is a very sensitive probe of the
gluon densities. In $pA$ fixed-target experiments, this process can be
used to study the gluonic analogue of the EMC effect. As we have shown in
this letter, a study of the ratio of the triple differential
cross-sections, integrated along a fixed \xtwo\ contour, for $pA$
and $pp$ collisions
will directly yield the ratio of gluon distributions in the nucleus and
the nucleon over a wide range of $x$. This information by itself  can be
used to discriminate between the different models of the EMC effect.
The power of discrimination can be increased further by combining this
with the measurement of this process at RHIC, which is sensitive to
\gA\ at smaller values of $x \sim 0.1 $. In this case it will not even be
necessary to measure the triple differential cross-section and a measurement
of $d \sigma / dp_T $ will suffice to distinguish between the different
models.

\newpage
\section*{Figure captions}
\renewcommand{\labelenumi}{Fig. \arabic{enumi}}
\begin{enumerate}
\item   
The predictions of the gas \cite{gas}, rescaling
\cite{rescale}, and quark-cluster model \cite{bag},
for the ratio of the nuclear gluon density \gA\ to that of the nucleon
\rhog , for $A = 56$, as a function of $x$, the momentum  fraction.  The
corresponding values of \pT\ for the FNAL fixed-target energy, $\sqrt{s} =
38.75$ GeV, are shown on the top axis.
\vspace{6mm}
\item  
The expected transverse momentum spectrum of the \jpsi\ and
$\gamma$ produced in the $pA\ (AA)$ collisions for the FNAL (RHIC), at
$\sqrt{s} = 38.75\ (100)$ GeV, for $A = 56$, for the gas, rescaling and
quark-cluster model. The spectrum is obtained by integrating over
$|y^{J/\psi},y^\gamma| \leq 2.0\ (4.0) $  for FNAL(RHIC) and the leptonic
branching ratio of the \jpsi\ into one lepton species has been folded in.
\vspace{6mm}
\item 
The triple differential cross-section $d\sigma/{dp_T\ dy^\gamma dy^{J/\psi}}$
(in pb/GeV), expected in the gas model,
for  $p + A \rightarrow J/\psi + \gamma \rightarrow l^+ l^- \gamma $, at the
fixed-target FNAL energy ($\sqrt{s} = 38.75$ GeV), as a function of  the
rapidities \yj\ and \ygam\ of the \jpsi\ and the $\gamma $ respectively,  at
$p_T = 3$ GeV and $A = 56$.
\vspace{6mm}
\item 
Contours of constant triple differential cross-section,
$d\sigma/{dp_T\ dy^\gamma dy^{J/\psi}}$,
for  $p + A \rightarrow J/\psi + \gamma \rightarrow l^+ l^- \gamma $, at
$\sqrt{s} = 38.75$ GeV, for $A = 56, p_T = 3$ GeV, for the gas (a),
rescaling (b) and quark-cluster model (c) of the nuclear parton densities.
The values shown are in pb/GeV.
Also superimposed are the contours of constant \xtwo, the momentum
fraction carried by the nuclear partons.
\end{enumerate}

\clearpage

\begin{thebibliography}{99}

\bibitem{emcdata} J.J.~Aubert et al., Phys. Lett. {\bf B 123}
(1983) 275.

\bibitem{emcrev} For a recent review see M.~Arneodo, Preprint,
CERN-PPE/92-113.

\bibitem{adep}F.~Close, R.~Roberts and G.G.~Ross, Phys. Lett.
{\bf 129} (1983) 346; A.~Harindranath and J.P.~Vary, Phys. Rev.
{\bf D 34} (1986) 3378; G.~Ballochi and M.~Zielinski, Z.~Phys
{\bf C 32} (1986) 27; R.M.~Godbole and S.~Gupta, Phys. Lett.
{\bf B 228} (1989) 129; R.M.~Godbole, in `` Frontiers in Particle Physics  '',
eds. Z.~Ajduk, S.~Pokorski and A.K.~Wroblewski, World Scientific,
Singapore (1990), p. 483; R.M.~Godbole and Sridhar K., Z.~Phys.
{\bf C 51} (1991) 417;
K.E.~Lassila and U.P.~Sukhatme, Z. Phys. {\bf C 44} (1991) 1188;
R.V.~Gavai and S.~Gupta, Z. Phys. {\bf C 49} (1991) 663.

\bibitem{sousri} S.~Gupta and Sridhar~K., Phys. Lett. {\bf B
197} (1987) 259; Sridhar~K., Z. Phys. {\bf C 55} (1992) 401.

\bibitem{nmc} D.~Allasia et al., Phys. Lett. {\bf B 258}
(1991) 493; P.~Amaudruz et al., Nucl. Phys. {\bf B 371}
(1992) 553.

\bibitem{csmod}E.L.~Berger and D.~Jones, Phys Rev. {\bf D 23}
(1981) 1521; R.~Baier and R.~R\"uckl, Nucl. Phys. {\bf B 201}
(1981) 1.

\bibitem{drekim}M. Drees and C.S. Kim, Z.~Phys {\bf C 53}
(1992) 673.

\bibitem{sri2}Sridhar K.,  Phys. Lett. {\bf B 289} (1992) 435.

\bibitem{satz} T.~Matsui and H.~Satz, Phys. Lett. {\bf B178} (1986) 416.

\bibitem{uaone}C.~Albajar et al., Phys. Lett. {\bf B 273} (1991)
540.

\bibitem{gas} S.~Gupta and K.V.L.~Sarma, Z. Phys. {\bf C 29} (1985) 329.

\bibitem{rescale} F.E.~Close, R.G.~Roberts and G.G.~Ross, Phys. Lett.
{\bf B 129} (1983) 346; S.~Gupta, B.~Banerjee and R.M.~Godbole, Z. Phys.
{\bf C 28} (1985) 483; S.~Gupta, Pramana {\bf 24} (1985) 443.

\bibitem{bag} U.~Sukhatme, G.~Wilk and K.E.~Lassila, Z. Phys. {\bf C 53}
(1992) 439.

\end{thebibliography}
\end{document}